\documentclass[11pt]{article} 
\newcommand{\ig}{\i\newcommand{\ee}{\end{equation}}ncludegraphics} 
\newcommand{\be}{\begin{equation}}
\newcommand{\ee}{\end{equation}}
\newcommand{\bav}{\begin{array}{c}}
\newcommand{\eav}{\end{array}}
\newcommand{\bam}{\begin{array}{cl}}
\newcommand{\eam}{\end{array}}
\newcommand{\noi}{\ensuremath{\nu_{i1}}}
\newcommand{\noj}{\ensuremath{\nu_{j1}}}
\newcommand{\nosi}{\ensuremath{\nu_{i1}^{\ast}}}

\newcommand{\nti}{\ensuremath{\nu_{i2}}}
\newcommand{\ntj}{\ensuremath{\nu_{j2}}}
\newcommand{\ntsi}{\ensuremath{\nu_{i2}^{\ast}}}
\newcommand{\ntsj}{\ensuremath{\nu_{j2}^{\ast}}}
\newcommand{\Noi}{\ensuremath{N_{i1}}}
\newcommand{\Noj}{\ensuremath{N_{j1}}}
\newcommand{\Nosi}{\ensuremath{N_{i1}^{\ast}}}
\newcommand{\Nosj}{\ensuremath{N_{j1}^{\ast}}}
\newcommand{\Nti}{\ensuremath{N_{i2}}}

\newcommand{\Ntsi}{\ensuremath{N_{i2}^{\ast}}}
\newcommand{\Ntsj}{\ensuremath{N_{j2}^{\ast}}}
\newcommand{\adot}{\dot{a}}
\newcommand{\bdot}{\dot{b}}
\newcommand{\half}{\frac{1}2}
\newcommand{\nubar}{\bar{\nu}}
\newcommand{\Nbar}{\bar{N}}
\newcommand{\sigmabar}{\bar{\sigma}}
\newcommand{\gbar}{\bar{g}} 
\newcommand{\SU}{\mathrm{SU}}

\newcommand{\U}{\mathcal{U}}

\newcommand{\M}{\mathcal{M}}
\newcommand{\mS}{\mathcal{S}}
\newcommand{\rH}{H_0^{\mathrm{red}}}
\newcommand{\Ho}{{\textstyle H_0}}
\newcommand{\one}{{\bf 1}}

\title{Neutrino Masses and Mixing with General Mass Matrices}
\author{Glenn D. Starkman  and  Dejan Stojkovic \\
Dept. of Physics, Case Western Reserve University, Cleveland, OH 44106}

\begin{document}
\maketitle
\abstract{We consider the most general neutrino masses and mixings
including Dirac mass terms, $M_D$,
as well as Majorana masses, $M_R$ and $M_L$.
Neither  the Majorana nor the Dirac mass matrices are 
expected to be diagonal in the eigenbasis of weak interactions,
and so the resulting eigenstates of the Hamiltonian
are admixtures of $\SU(2)_L$ singlet and doublet
fields of different ``generations.'' 
We show that for three generations each of doublet and singlet neutrinos, 
diagonalization of the Hamiltonian to obtain the propagating
eigenstates in the general case requires diagonalization of a
$12\times12$ Hermitian matrix, 
rather than the traditional $6\times6$ symmetric mass matrix.
The symmetries of the $12\times12$  matrix {\em are} such that 
it has $6$  pairs of real eigenvalues.
Although the standard ``see-saw" mechanism remains valid, 
and indeed the eigenvalues obtained are identical to the
standard ones, the correct description
of diagonalization and mixing is more complicated.
The analogs of the CKM matrix
for the light and the heavy neutrinos are nonunitary, enriching the  
opportunities for CP violation  in the full neutrino sector.}

\section{Introduction}

One of the most intriguing features of the Standard Model  
is the complex mismatch between 
quark eigenstates of the weak interactions and 
the freely propagating eigenstates. Particularly
intriguing is the resulting   CP violation in the 
weak charged current interactions.
In the Standard Model, this mismatch is confined to the quark sector,
and by convention to the down-type quarks.
The charged leptons, like the up-type quarks, 
can be chosen to have identical mass and weak eigenstates. 
Meanwhile, the absence of $\SU(2)_L$ singlet neutrino fields  
precludes a Dirac mass for the neutrino,
while a purely Majorana mass for the $\SU(2)_L$ doublet neutrinos 
seems to be in conflict with experimental results 
on the width of the Z boson and the $\rho$-parameter.
This eliminates the possibility of any neutrino masses or mixings,
and any of the associated CP violation, unless new fields are added
to the theory.

It has long been suspected that neutrinos are not absolutely massless.  
Many GUT theories and left-right symmetric theories 
imply the existence of $SU(2)_L \times U(1)_Y$ 
singlet fields, often called sterile neutrinos, 
and hence the possibility of non-zero masses for both sterile 
and active ($SU(2)_L$ doublet) neutrinos.
Over the past thirty years, 
considerable experimental evidence has
accumulated to support the  view that neutrinos are not massless.
These include the large deficit in  neutrinos 
arriving from the Sun compared to the predictions of the Standard Model
of particle physics together with standard solar models.
This deficit, it has been shown, cannot be explained by
changes in the solar model.  
The only fully self-consistent explanation 
for all the solar neutrino data is that  
the electron neutrinos emitted in the Sun  oscillate
into other neutrino states 
with smaller cross-sections for detection.
(For a recent review of the problem see \cite{Bahcall} 
and references therein.)
Such oscillations require non-zero neutrino masses and mixings.
Similarly in the flux of neutrinos 
produced by cosmic rays in the atmosphere 
there is a  deficit in the observed ratio 
of muon neutrinos to electron neutrinos 
compared to  the predicted ratio \cite{SuperK},
as well as an up-down asymmetry in the flux of muon neutrinos.
These can be explained by the oscillation of muon neutrinos  
into tau or sterile neutrinos.
The null hypothesis of no neutrino mixing 
is strongly rejected  by both the 
Kamiokande and Super-Kamiokande data \cite{SuperK} 
which also are consistent with the preferred parameters for an 
explanation in terms of neutrino oscillations.
One laboratory experiment, the LSND \cite{LSND},
also finds evidence for neutrino oscillations.

Given the overwhelming experimental evidence 
and theoretical motivation for neutrino masses and mixings, 
a full and careful treatment of the problem is warranted.  
Previous discussions have often focused, sometimes deliberately,
on special cases, in particular on a real neutrino mass matrix.
Although the masses obtained by the usual ``see-saw" mechanism \cite{ss} 
remain unchanged in the full theory, 
the detailed mechanics of treating general mass models are more complicated. 
In particular, diagonalization of the Hamiltonian is not equivalent to
diagonalization of the standard mass matrix. Also,
the leptonic CKM matrix is {\em not} a $3\times3$ unitary matrix
in generation space, but rather two $3\times3$ nonunitary matrices,
one for the light and one for the heavy mass eigenstates 
(see section \ref{sec:CP}).

\section{Mass eigenstates in neutrino theory}

Suppose that the three $\SU(2)_L$ doublet neutrinos of the Standard 
Model, $\nu_i$ ($i=e,\mu,\tau$), are supplemented by three $\SU(2)_L$ 
singlet neutrinos $N_i$. 
$\nu_{ia}$ and $N_{ia}$ are two-component Weyl fermions,
with definite Lorentz properties --
the index $a$ signifying that they  
transform under the $({1\over2},0)$ representation.
(In principle one can have more or less than three singlet neutrinos and
this treatment generalizes in an obvious way.)
The general free-field Lagrangian density for the neutrinos is

\begin{eqnarray}
\label{2cL}
-{\cal L} 
= & i N_{ia} \sigma^{\mu a \bdot} \partial_{\mu} \Nbar_{i\bdot} 
 + i \nubar_{i\adot} \sigmabar^{\mu \adot b} \partial_{\mu} \nu_{ib}
 + \nubar_{i\adot} \, M_{Dij} \, \Nbar_j^{\adot} 
 + N^{a}_{i} \, M^{\dagger}_{Dij} \, \nu_{ja} \\
&+ \half \, \Nbar_{i\adot} \, M_{Rij} \, \Nbar^{\adot}_{j} 
+\half \, N^{a}_{i} \, M^{\dagger}_{Rij} N_{ja}  
+ \half \, \nubar_{i\adot} \, M_{Lij} \, \nubar_{j}^{\adot} 
+\half \, \nu_{i}^{a} \, M^{\dagger}_{Lij} \nu_{ja}  \nonumber 
\end{eqnarray}
\noindent
Greek letters ($\mu=0,1,2,3$) denote Lorentz four-vector indices,
with $\sigma^{\mu}_{a \dot{b}} = (\one, \vec{\sigma})_{a \dot{b}}$ and 
$\sigmabar^{\mu \dot{a}b} =  (\one, -\vec{\sigma})^{\dot{a}b} $.
($ \vec{\sigma}$ are the usual Pauli matrices.)
Latin letters denote either Weyl spinor  indices ($a,b=1,2$)
or generation indices ($i,j=1,2,3$).
The dotted spinor indices ($\adot,\bdot$) attached to $\nubar_i$ and $\Nbar_i$ 
indicate that the complex conjugates $\nu^{\ast}_i$ and $N^{\ast}_i$
transform under the $(0,{1\over 2})$ representation of the Lorentz group ---
\noindent $(\nu_{a})^{\ast} = \nubar_{\adot}$ \ and \ 
$(N^{a})^\ast=\Nbar^{\adot}$.
Spinor indices, dotted and undotted, are raised and lowered by 
$g_{ab}$ and $\gbar^{\adot \bdot}$ which act
as metric tensors in spinor space :
\be
 g_{ab}  =  i (\sigma^2)_{ab} = \varepsilon_{ab}  \quad {\rm and} \ 
 \gbar^{\adot \bdot} =  i (\sigmabar^2)^{\adot\bdot} =
-\varepsilon^{\adot\bdot} 
\ee
with 
$\varepsilon_{12} = \varepsilon_{\dot{1}\dot2} 
= \varepsilon^{12} = \varepsilon^{\dot{1}\dot2} = 1 $.
The inverses of $g_{ab}$ and $\gbar^{\adot \bdot}$ are respectively
$g^{ab} = -\varepsilon^{ab}$ 
and $\gbar_{\adot \bdot}= \varepsilon_{\adot\bdot}$.
Thus, $\sigma^{\mu a \dot{b}} = (\one, -\vec{\sigma}^{\ast})^{a \dot{b}}$ 
and $(\sigma^{\mu a \dot{b}})^{\ast} = (\one, -\vec{\sigma})^{\dot{a}b} \equiv 
\bar{\sigma}^{\mu \dot{a}b}$ 

$M_D$ , $M_L$ and $M_R$ are complex matrices.
$M_R$ and $M_L$  are symmetric, since
$\Nbar_{i\dot{a}} M_{Rij}\Nbar_{j}^{\adot} 
=-\Nbar_j^{\adot} M_{Rij}\Nbar_{i\adot}
=+\Nbar_{j\adot} M_{Rij}\Nbar_{i}^{\adot}$,
and likewise for $M_L$.

It is convenient to make explicit in ${\cal L}$ 
the dependence on spinor indices $a={1,2}$. 
So long as no freely propagating eigenstates are massless,
we can, without loss of generality, do this in the rest frame
in which ${\vec\partial} N_{ia} = {\vec\partial} \Nbar_i^{\adot} =
{\vec\partial} \nu_{ia} = {\vec\partial} \nubar_i^{\adot} = 0 $.
This will allow us to find the zero-momentum eigenmodes of the system.
Neutrinos of non-zero momentum are related to these 
by appropriate Lorentz boosts.
\begin{eqnarray}
\label{Lcomp}
 -{\cal L} \!\!\!\!\!\!&= i \Noi \partial_{0} \Nosi \!+\! i \Nti \partial_{0} \Ntsi 
\!+\! i \nosi \partial_{0} \noi \!+\! i \ntsi \partial_{0} \nti 
\! - \!\nosi M_{Dij} \Ntsj \!+\! \ntsi M_{Dij} \Nosj \\
&\!\!-\! \Nti M^{\dagger}_{Dij} \noj \!\!+\!\! \Noi M^{\dagger}_{Dij} \ntj 
\!\!-\!\!\Nosi M_{Rij} \Ntsj \!\!-\!\! \Nti M^{\dagger}_{Rij} \Noj 
\!\!-\!\!\nosi M_{Lij} \ntsj \!\!-\!\! \nti M^{\dagger}_{Lij} \noj \nonumber 
\end{eqnarray}
\noindent

The equations of motion that follow from the Lagrangian density 
(\ref{Lcomp}) are (suppressing generation indices):

\be
\label{eom}
\frac{i}{2}\frac{\partial}{\partial t} 
\underbrace{\left(\!\! \bav \nu_1 \\ N_1 \\ \nu^{\ast}_2 \\ N^{\ast}_2 \\
 \nu_1^{\ast} \\ N^{\ast }_1 \\ \nu_2 \\ N_2 \eav\!\! \right)}_{\textstyle \Psi} =
\underbrace{ \left( \begin{array}{cccccccc}
0&\!\!0&\!\!M_L&\!\!M_D&\!\!0&\!\!0&\!\!0&\!\!0 \\ 
0&\!\!0&\!\!M^T_D&\!\!M_R&\!\!0&\!\!0&\!\!0&\!\!0 \\
M^{\dagger}_L&\!\!M^{\ast}_D&\!\!0&\!\!0&\!\!0&\!\!0&\!\!0&\!\!0 \\
M^{\dagger}_D&\!\!M^{\dagger}_R&\!\!0&\!\!0&\!\!0&\!\!0&\!\!0&\!\!0 \\
0&\!\!0&\!\!0&\!\!0&\!\!0&\!\!0&\!\!-M^{\dagger}_L&\!\!-M^{\ast}_D \\ 
0&\!\!0&\!\!0&\!\!0&\!\!0&\!\!0&\!\!-M^{\dagger}_D&\!\!-M^{\dagger}_R \\
0&\!\!0&\!\!0&\!\!0&\!\!-M_L&\!\!-M_D&\!\!0&\!\!0 \\
0&\!\!0&\!\!0&\!\!0&\!\!-M^T_D&\!\!-M_R&\!\!0&\!\!0
 \end{array} \right) }_{\Ho} 
\underbrace{\left(\!\! \bav \nu_1 \\ N_1 \\ \nu^{\ast}_2 \\ N^{\ast}_2 \\
 \nu_1^{\ast} \\ N^{\ast }_1 \\ \nu_2 \\ N_2 \eav\!\! \right)}_{\textstyle \Psi} 
\ee
\noindent
$M_D$, $M_R$ and $M_L$ should be interpreted as $3\times3$ matrices
in generation space, while $\nu_a$, $\nu_a^\ast$, $N_a$ and $N_a^\ast$ ($a=1,2$)
are each three-dimensional vectors in generation space.
$\Ho$ is the free, zero-momentum Hamiltonian operator of the theory. 
Its eigenstates are the zero-momentum normal modes ---
free-field propagating degrees of freedom in the theory.
$\Ho$ is also the ``mass" matrix in the same basis, {\it i.e.}
\be
\mathcal{L}_{mass} =\Psi^{\dagger} \, H_0 \, \Psi 
\ee
\noindent

This ensures that the zero-momentum 
free eigenmodes are in fact mass eigenstates, as would be expected.
The eigenvalues are the masses of the physical neutrinos. However, it is crucial
to note that $H_0$ is not the standard mass matrix,
$\M = \left( \bam M_L&M_D \\ M_D^{T}&M_R \eam \right)$.
$H_0$ is defined in both spinor and generation space, while $\M$ acts only in
generation space.

Fortunately, we are not required to manipulate
the monolithic $24\times24$ $\Ho$, 
since 
\be
H_0= \left( \bam \rH &0\\0&-(\rH)^{\ast} \eam \right).
\ee
The matrix $\rH$ is Hermitian, and the complex conjugate  of the
unitary transformation which 
diagonalizes $\rH$ will diagonalize $(\rH)^{\ast}$. 
$\rH$ acts on 
\be
\Psi^{\mathrm{red}}\equiv
\left( \bav \nu_1 \\ N_1 \\ \nu^{\ast}_2 \\N^{\ast}_2 \eav \right)
\ee
while $H_0$ acts on 
$\Psi \equiv \left( \Psi^{\mathrm{red}} 
\ (\Psi^{\mathrm{red}})^{\ast} \right)^T$. 
The complete diagonalization procedure 
will be presented in section (\ref{sec:General}).

\section{The Case of One Generation}
\label{sec:COG}

In the case of one generation the matrix
which we must diagonalize is 
\be
\left( \begin{array}{cccc}
0& 0& m_L e^{i \theta}& m_D e^{i\varphi} \\
0& 0& m_D e^{i\varphi}& m_R e^{i\phi}\\
m_L e^{-i \theta}& m_D e^{-i\varphi} &0&0 \\
m_D e^{-i\varphi}&m_R e^{-i\phi}&0&0 
\end{array} \right)
\ee
For the sake of simplicity, we take $m_L=0$.
The eigenvalues of this matrix are $(m_+,m_-,-m_-,-m_+)$ with 
\be
m_{\pm} = \half (\sqrt{4m_D^2+m_R^2} \pm m_R)
\ee
For $m_D\ll m_R$,
$m_+ \simeq m_R$, while $m_- \simeq m_D^2/m_R$,
the usual ``see-saw" mechanism results \cite{ss}.
The corresponding eigenvectors are:
\be
\label{ev11}
\left( \bav
\frac{e^{i(\varphi -\frac{\phi}{2})}m_-}{M_D} \\  e^{i \frac{\phi}{2}} \\ 
\frac{e^{i(\frac{\phi}{2}-\varphi)}m_-}{M_D}
\\ e^{-i\frac{\phi}{2}}  \eav \right) \ \ \ \ 
\left( \bav  \frac{e^{i(\varphi -\frac{\phi}{2})}m_+}{M_D} \\ - e^{i \frac{\phi}{2}} \\ 
\frac{-e^{i(\frac{\phi}{2}-\varphi)}m_+}{M_D} \\ e^{-i\frac{\phi}{2}} \eav \right)  \ \ \ \ 
\left( \bav - \frac{e^{i(\varphi -\frac{\phi}{2})}m_+}{M_D} \\  e^{i \frac{\phi}{2}}\\ 
\frac{-e^{i(\frac{\phi}{2}-\varphi)}m_+}{M_D} \\ e^{-i\frac{\phi}{2}} \eav \right)  \ \ \ \
\left(\bav -\frac{e^{i(\varphi -\frac{\phi}{2})}m_-}{M_D} \\ - e^{i \frac{\phi}{2}} \\ 
\frac{e^{i(\frac{\phi}{2}-\varphi)}m_-}{M_D} \\ e^{-i\frac{\phi}{2}} \eav \right) 
\ee

Some interesting properties of these eigenvectors
will be discussed in section (\ref{sec:General}) and (\ref{sec:decoupling}).

\section{The General Case}
\label{sec:General}

In general $M_D \neq 0$, $M_R \neq 0$, $M_L \neq 0$, 
and there are three generations of doublet neutrinos
(although in many extensions of the standard model
$M_L=0$ identically).  
As discussed above,
there can be any number of singlet neutrinos, 
but the Dirac mass terms will couple 
only three independent linear combinations of
the singlet neutrinos to the doublet neutrinos.
If there are fewer than three singlet neutrinos,
then this is equivalent to appropriate zero entries
in $M_D$, $M_R$ and $M_L$. 
We will therefore assume that there are three singlet neutrino species.
This leaves us to diagonalize
a higher dimensional matrix which contains $M_D$, $M_R$ and $M_L$. 
Let us, as is often done, introduce the new fields:
\be \hspace{-5mm}
f_\pm \equiv \frac{\psi_L \pm (\psi_L)^c}{\sqrt2} , \ \  {\rm and} \ \ \ 
F_\pm \equiv \frac{\psi_R \pm (\psi_R)^c}{\sqrt2} 
\ee
and rewrite ${\cal L}$ as:
\begin{eqnarray}
\label{eqn:Lplusminus}
 -{\cal L} = & \half i\left(\overline{f_+} \gamma^{\mu} \partial_{\mu} f_+ + 
\overline{F_+} \gamma^{\mu} \partial_{\mu} F_+ + \overline{f_-} \gamma^{\mu} \partial_{\mu}
f_- + \overline{F_-} \gamma^{\mu} \partial_{\mu} F_-\right) +  \\
+ \half \left( \overline{f_+} \ \overline{F_+} \ \overline{f_-} \ \overline{F_-} \right) 
& \left( \begin{array}{cccc} 
M_L+M^{\dagger}_L & M_D + M^{\ast}_D & M_L-M^{\dagger}_L & M_D - M^{\ast}_D \\
M_D^{T}+M^{\dagger}_D & M_R+M^{\dagger}_R & M^{\dagger}_D-M^T_D & M_R-M^{\dagger}_R \\ 
M^{\dagger}_L-M_L & M_D-M^{\ast}_D & -M_L-M^{\dagger}_L & M_D+M^{\ast}_D \\
M_D^{\dagger}-M^T_D & M^{\dagger}_R-M_R & M^T_D+M^{\dagger}_D & -M_R-M^{\dagger}_R      
\end{array} \right) 
\left( \bav f_+ \\ F_+ \\ f_- \\ F_- \eam \right) \nonumber
\end{eqnarray}

$M_D$, $M_R$ and $M_L$ are $3\times3$ matrices in generation space.
If all three of these matrices are real, 
then the cross terms between $+$ fields and $-$ fields vanish 
and the theory decouples into two sectors: ``+'' and ``-".
\begin{eqnarray}
\label{ss}
 -{\cal L} = & \half i\overline{f_+} \gamma^{\mu} \partial_{\mu} f_+ + 
\half i\overline{F_+} \gamma^{\mu} \partial_{\mu} F_+ + \half \left( \overline{f_+} \ \overline{F_+} \right)
\underbrace{ \left( \bam M_L&M_D \\ M_D^{T}&M_R \eam \right) }_{\textstyle{\cal M}_+}
\left( \bav f_+ \\ F_+ \eam \right) + \\ +
 & \half i\overline{f_-} \gamma^{\mu} \partial_{\mu} f_- + 
\half i\overline{F_-} \gamma^{\mu} \partial_{\mu} F_- + \half \left( \overline{f_-} \ \overline{F_-} \right)
\underbrace{ \left( \bam -M_L&M_D \\ M_D^{T}&-M_R \eam \right)}_{\textstyle{\cal M}_-}
\left( \bav f_- \\ F_- \eam \right) \nonumber
\end{eqnarray}

The matrix $\M_+ \equiv \M$ is just a standard ``see-saw" matrix. 
$\M_-$ has eigenvalues with equal magnitudes but
opposite signs from the eigenvalues of $\M_+$.
Making the change of variables 
$f_- \rightarrow -\gamma_5 f_-$ and $F_- \rightarrow \gamma_5 F_-$, 
which keeps kinetic terms invariant, we convert $\M_-$ to $\M_+$. 
The theory is thereby rewritten in terms of Majorana fields, 
$(f_{\pm})^c=f_{\pm}$ and $(F_{\pm})^c=F_{\pm}$, 
and the eigenstates of the mass matrices are Majorana fields as well.

We can not, or course, confine ourselves 
just to the ``+'' sector or just to the ``-" sector, 
even though these two sectors have identical structures.
Doing that we would lose half of original degrees of freedom. 
Also, eigenstates in these two sectors can have different properties 
(like helicity, P-parity, CP-parity, ...).

If the mass parameters are complex, 
then the cross terms in equation (\ref{eqn:Lplusminus})
do not vanish and the situation is more complicated. 
To find the eigenstates we must get rid of cross terms, 
{\it i.e.} diagonalize the theory.

We continue discussion in the basis 
where the connection to the two-component formalism is obvious.
\begin{eqnarray}
-{\cal L} 
= i \overline{\psi_L} \gamma^{\mu} \partial_{\mu} \psi_L 
+ i \overline{\psi_R} \gamma^{\mu} \partial_{\mu} \psi_R \ +& \\
 + \half (\overline{\psi_L} \ \overline{(\psi_R)^c})
\left(\!\!\!\!\bam M_L &\!\!M_D \\ M_D^{T} &\!\!M_R \eam\!\!\!\!\right)&
\!\!\!\!\left(\!\!\!\!\bav (\psi_L)^c \\ \psi_R \eav\!\!\!\!\right)   +
\half (\overline{(\psi_L)^c} \ \overline{\psi_R})
\left(\!\!\!\!\bam M^{\dagger}_L &\!\! M^{\ast}_D \\ M^{\dagger}_D &\!\! M^{\dagger}_R \eam\!\!\!\!
 \right)
\left(\!\!\!\!\bav \psi_L \\ (\psi_R)^c \eav\!\!\!\!\right)\nonumber   \cr
\end{eqnarray}

\noindent where we used $\overline{\psi_L} \psi_R = \overline{(\psi_R)^c}
(\psi_L)^c $. In more compact notation $\mathcal{L}_{mass}$ is:
\be
\label{m}
-{\cal L}_{mass} = 
\half (\overline{(n_R)^c}\ \overline{n_R} )
\left( \begin{array}{cc} 0 & \M \\ \M^{\dagger} & 0
\end{array} \right)
\left( \bav  (n_R)^c \\ n_R \eav \right)
\ee
\noindent where $n_R= \left( (\psi_L)^c \ \ \psi_R \right)^T$.
The mass matrix in (\ref{m}) is just $\rH$.

The first step in diagonalizing $\rH$ 
is a unitary transformation with the matrix 
\be
S^{{\mathrm red}} \equiv  \left( \begin{array}{cc} \U^T & 0 \\ 
0 & \U^{\dagger} \end{array} \right) ,
\ee
where $\U$ diagonalize $\M$: 
$\U^T \M \U = \M_d$ with $\M_d$ a real, positive, diagonal matrix.
 So, we have:
\be \label{stand}
S^{\mathrm{red}}\, \rH \, (S^{\mathrm{red}})^{\dagger} = 
\left( \bam 0 & \M_d \\ \M_d & 0 \eam \right) 
\ee

It is worth pointing out that the standard treatment stops here. If we neglect
spinor degrees of freedom, matrix (\ref{stand}) is diagonal in generation space.
But, clearly, the diagonalization is not yet complete and we still do not have
the propagating degrees of freedom --- the normal modes of the theory.

We convert matrix (\ref{stand}) into diagonal form by an extra unitary 
transformation with the matrix \hfil\break
$\mathcal{X} \equiv 
\frac{1}{\sqrt{2}} \left( \bam 1 & 1 \\ \! \! -1 & 1 \eam \right)$:
\be
\mathcal{X}  \left( \bam 0 & \M_d \\ \M_d & 0 \eam \right) \mathcal{X}^{\dagger}
= \left( \bam \M_d & 0 \\ 0 & -\M_d \eam \right)
\ee
The complete transformation which diagonalizes $\rH$ is thus
$\mS \equiv \mathcal{X}
S^{\mathrm{red}}$. After diagonalization, we can write:
\be
-\mathcal{L}_{mass}=\half 
%\left(\overline{(n_R)^c}U^{\ast} + \overline{n_R} U , \ 
%-\overline{(n_R)^c}U^{\ast} + \overline{n_R} U\right) 
\left({\bar\chi_+}, {\bar \chi}_-'\right)
\left( \bam \M_d & 0 \\ 0 & -\M_d \eam \right) 
\left(\bav \chi_+ \\ \chi_-'\eav\right)
%\left( \bav U^T (n_R)^c + 
%U^{\dagger} n_R \\ -U^T (n_R)^c + U^{\dagger} n_R \eav \right)
\ee
or
\be
\label{semifinal}
-\mathcal{L}_{mass}=\half \overline{\chi_+}\M_d \chi_+ - \half \overline{\chi_-}^{'}\M_d \chi^{'}_-
\ee
\noindent where $\chi_+ \equiv U^T (n_R)^c + U^{\dagger} n_R$ and
$\chi^{'}_- \equiv -U^T (n_R)^c + U^{\dagger} n_R$.
With the redefinition $\chi_- \equiv \gamma_5 \chi^{'}_-$, $-\mathcal{L}_{mass}$
acquires its final diagonal form:

\be
\label{final}
-\mathcal{L}_{mass}= \half \overline{\chi_+} \M_d \chi_+  + 
\half \overline{\chi_-}\M_d\chi_-
\ee
(Note: Rosen \cite{Rosen} also saw the necessity of diagonalizing
the 12x12 Hamiltonian and arrived  at (\ref{semifinal}), 
but suggested that the $\chi^{'}_-$ were unphysical eigenstates
of negative mass.)

Restoring generation indices, we have:
\be \label{result}
-\mathcal{L}_{mass}= \half \sum_{k=1}^{2n} m_k \overline{\chi_+}_k \chi_{+k} 
 + \half \sum_{k=1}^{2n} m_k \overline{\chi_-}_k \chi_{-k}
\ee
\noindent where $m_k$ are the diagonal elements of $\M_d$,
%$\chi=\left( \chi_1 \ \chi_2 \ \hdots \right)^T$ 
and $n$ is the number of generations ($n=3$). 
$\chi_{\pm}$ are Majorana fields, i.e.
$\chi_{\pm}^c = \chi_{\pm}$, so, in the case of $n$ generations we have 
$4n$ Majorana neutrinos. In the original theory we had $2n$ complex degrees
of freedom, while the final theory has 4n real degrees of freedom. 
The particles $\chi_+$ and $\chi_-$ have the same masses
but may have different symmetry properties and 
possibly different interactions. We should point out that result 
(\ref{result}) differs from the standard results found in much of the
literature.

In section (\ref{sec:COG}) we explicitly derived 
the eigenvectors of the mass matrix in the one generation case. 
In terms of the weak eigenstates,
the propagating eigenvectors of the theory 
(which are just adjoints of the mass matrix eigenstates (\ref{ev11})) are:
\begin{eqnarray}
\chi_{+1}&=-\frac{e^{-i(\varphi -\frac{\phi}{2})}m_+}{M_D} \psi_L +
e^{-i \frac{\phi}{2}} (\psi_R)^c  
-\frac{e^{-i(\frac{\phi}{2}-\varphi)}m_+}{M_D} (\psi_L)^c +
e^{i\frac{\phi}{2}} \psi_R \\ \nonumber
\chi_{+2}&=\frac{e^{-i(\varphi -\frac{\phi}{2})}m_-}{M_D} \psi_L +
e^{-i \frac{\phi}{2}} (\psi_R)^c + 
\frac{e^{-i(\frac{\phi}{2}-\varphi)}m_-}{M_D} (\psi_L)^c +
e^{i\frac{\phi}{2}} \psi_R \\ \nonumber
\chi^{'}_{-1}&=\frac{e^{-i(\varphi -\frac{\phi}{2})}m_+}{M_D} \psi_L 
-e^{-i\frac{\phi}{2}} (\psi_R)^c 
-\frac{e^{-i(\frac{\phi}{2}-\varphi)}m_+}{M_D} (\psi_L)^c +
e^{i\frac{\phi}{2}} \psi_R \\ \nonumber
\chi^{'}_{-2}&=-\frac{e^{-i(\varphi -\frac{\phi}{2})}m_-}{M_D} \psi_L 
-e^{-i \frac{\phi}{2}} (\psi_R)^c + 
\frac{e^{-i(\frac{\phi}{2}-\varphi)}m_-}{M_D} (\psi_L)^c +
e^{i\frac{\phi}{2}} \psi_R 
\end{eqnarray}

We finally conclude that to obtain the propagating
eigenstates for the general case
we must diagonalize the $12\times 12$ Hermitian matrix 
\be
H^{{\mathrm red}} = \left( \begin{array}{cccc} 0&0&M_L&M_D \\ 0&0&M^T_D&M_R \\ 
M^{\dagger}_L&M^{\ast}_D&0&0 \\ M^{\dagger}_D&M^{\dagger}_R &0&0\end{array} \right)
\ee

A general $12\times 12$ complex Hermitian matrix  is diagonalized
by a $12\times 12$ unitary matrix.  Unitary matrices of this size
are parametrized by 66 angles and 78 phases.  However,
the large numbers of symmetries of the Hamiltonian in this 12 dimensional
representation reduce the number of actual angles
and phases significantly.  After all, $M_D$, $M_R$ and $M_L$
contain only $18$, $12$ and $12$ real Lagrangian parameters respectively.
Algebraic analysis of the Hamiltonian indicates that its characteristic
equation is of the form
\be
\Pi_{i=1,...,6,} (\lambda^2-m_i^2) = 0
\ee
with $m_i$ real. $m_i$ can be obtained by a suitable diagonalization 
of the  $6\times6$ symmetric submatrix
\be
{\cal M} = \left(\begin{array}{cc} M_L&M_D\\M_D^T&M_R \end{array}\right),
\ee
but $m_i$ are not the eigenvalues of ${\cal M}$ --- ${\cal M}U=U^\ast M_d$ and
not ${\cal M}U=U M_d$.

There are six distinct absolute values of the eigenvalues (we saw that 
all of them can be made positive).   The $6\times6$ unitary matrices
diagonalizing ${\cal M}$ are parametrized in terms of 
$15$ angles and $21$ phases.

\section{Decoupling}
\label{sec:decoupling}
Examination of the set of eigenvectors (\ref{ev11}) suggests at first sight
that the ``light" and ``heavy" degrees of freedom 
may not be completely decoupled in the limit of infinitely large $M_{R}$. 
One can see this by inspection of one of the light
eigenvectors (an eigenvector which corresponds to a light eigenvalue). 
The presence of $\phi$, 
which is the phase of a Majorana mass parameter, 
even after we take limit of large $M_{R}$, is the origin of this concern. 

In three generations,
it is not possible to absorb all $6$ phases from $M_R$ by phase rotations
on $N_{i1}$ and $N_{i2}$. Does this imply a non-decoupling theorem? No.

Consider the relevant large $M_R$ limit $M_R/M_D\gg1$. 
Let $S$ be the $12\times 12$ matrix  whose columns  consist
of the eigenvectors of $M^{{\mathrm red}}$. 
$S$ is a unitary matrix, which we can write in the form
\be
S = \left(\begin{array}{cc} S_{\ell a} & S_{\ell s} \\ S_{H a} & S_{H s}
\end{array}\right)
\ee
where the subscripts $a$ and $s$ refer to the ``active'' (doublet) and ``sterile''
(singlet) states, 
and 
the subscripts $\ell$ and $H$ refer to the light ($m={\cal O}\left({M_D^2\over M_R}\right)$)
and heavy ($m={\cal O}\left(M_R\right)$) states.
Since $S_{\ell a}$ and $S_{Hs}$ are ${\cal O}(1)$, 
while $S_{\ell s}$ and $S_{Ha}$ are ${\cal O}\left({M_D\over M_R}\right)$,
the unitarity of $S$ implies that $S_{\ell a}$ 
is unitary at ${\cal O}\left({M_D^2\over M_R^2}\right)$.
Thus, in the limit that $M_R\to\infty$ when all the light states are 
massless and degenerate, we can make a unitary transformation
(at ${\cal O}\left({M_D^2\over M_R^2}\right)$ to the basis where
the active weak eigenstates are the light mass eigenstates, with
no remaining phases. This seems to indicate that all
physical processes exploring the mixing between active and sterile
states {\em are} suppressed by  powers of $M_D/M_R$.  

\section{Mixing and CP Violation}
\label{sec:CP}

The full consequences of a general complex neutrino mass matrix
for CP violation in the leptonic sector are currently 
under investigation.   The general mass matrix contains 
21 magnitudes and 21 phases (since $M_D$ is a general complex
matrix, while $M_L$ and $M_R$ are symmetric).  Of these, perhaps 6 phases
can be removed by phase redefinitions of the neutrino fields, leaving 15.
How many of these are physical? What is the subgroup of $U(6)$ from
which the unitary matrix diagonalizing the $6\times6$ submatrix  ${\cal M}$
should be drawn? We reserve most details for future publications.  

In particular, let us express the charged current, 
$J^+ =\overline{\psi_{iL}}\gamma^{\mu} l_{iL} $ \ 
($\psi_{i} = \nu_{e} , \nu_{\mu} , \nu_{\tau}$;
\ \   $l_i=e,\mu, \tau $)
in terms of the weak eigenstates:

\begin{eqnarray}
J^+ =\overline{\psi_{iL}}\gamma^{\mu} l_{iL}  = &\\ 
&=\underbrace{\left( \overline{\psi_{iL}} \ \overline{(\psi_{iR})^c} \ 
\overline{(\psi_{iL})^c} \ \overline{\psi_{iR}} \right)}
_{\overline{\textstyle{\Psi_W^{(\nu)}}}}
\underbrace{ \left( \begin{array}{cccc} \gamma^{\mu} &0&0&0\\
0&0&0&0\\0&0&0&0\\0&0&0&0 \end{array} \right) \otimes \bf{I_3}}_{\textstyle{G}} 
\underbrace{ \left( \begin{array}{c} 
l_{iL} \\ (l_{iR})^c \\ (l_{iL})^c \\ l_{iR} \end{array} \right)}_
{\textstyle{\Psi_W^{(l)}}} \nonumber
\end{eqnarray}

\noindent where the subscript ``W" 
denotes a weak eigenstate and $\bf{I_3}$
is three-dimensional unit matrix in generation space.
The matrix $\mS =\mathcal{X} S^{\mathrm{red}}
=\left( \bam \U^T & \U^{\dagger} \\ -\U^T & \U^{\dagger} \eam \right)$ 
brings the weak eigenbasis into the mass eigenbasis and we have:
\be
J^+ = \overline{\Psi^{(\nu)}_W} \mS^{(\nu) \dagger} \mS^{(\nu)} G \mS^{(l) \dagger}
\mS^{(l)} \Psi^{(l)}_W =
\overline{\Psi^{(\nu)}_M} \mS^{(\nu)} G \mS^{(l) \dagger} \Psi^{(l)}_M .
\ee
(We have identified separate transformations matrices 
$\mS^{(\nu)}$ and  $\mS^{(l)}$ for the neutrinos 
and for the charged leptons.)

Decompose the matrices  $\U^{(\nu)}$ and $\U^{(l)}$
into four block elements each:
\be
\U^{(\alpha)}=\left( \begin{array}{cc}
\U^{(\alpha)}_{11}&\U^{(\alpha)}_{12}\\\U^{(\alpha)}_{21}&\U^{(\alpha)}_{22} 
\end{array} \right) ,
\ee 
where block elements are matrices in generation space
(i.e. $\U^{(\alpha)}_{11} \equiv (\U_{11})_{ij}$ \ldots).
Using the fact that 
$\U^{(l)}$ has block diagonal form ($\U^{(l)}_{12}=\U^{(l)}_{21}=0$),
we can write:
\be
J^+= \left(\overline{\chi^{(\nu)}_{+1i}} \ \overline{\chi^{(\nu)}_{+2i}} \ 
\overline{\chi^{(\nu)'}_{-1i}} \ \overline{\chi^{(\nu)'}_{-2i}} \right)
  \left( \begin{array}{cccc} 
[(\U^{\nu}_{11})^T \gamma^{\mu} (\U^{l}_{11})^{\ast} ]_{ij} &0&
-[(\U^{\nu}_{11})^T \gamma^{\mu} (\U^{l}_{11})^{\ast} ]_{ij}&0 \\ 
\! [(\U^{\nu}_{12})^T \gamma^{\mu} (\U^{l}_{11})^{\ast} ]_{ij}&0&
-[(\U^{\nu}_{12})^T \gamma^{\mu} (\U^{l}_{11})^{\ast} ]_{ij}&0 \\
\! -[(\U^{\nu}_{11})^T \gamma^{\mu} (\U^{l}_{11})^{\ast} ]_{ij}&0&
[(\U^{\nu}_{11})^T \gamma^{\mu} (\U^{l}_{11})^{\ast} ]_{ij}&0 \\
\!- [(\U^{\nu}_{12})^T \gamma^{\mu} (\U^{l}_{11})^{\ast} ]_{ij}&0&
[(\U^{\nu}_{12})^T \gamma^{\mu} (\U^{l}_{11})^{\ast} ]_{ij}&0
\end{array} \right)
\left( \bav \chi^{(l)}_{+1j} \\ \chi^{(l)}_{+2j} \\ \chi^{(l)'}_{-1j} 
\\ \chi^{(l)'}_{-2j} \eav \right)
\ee

According to this, we can write analogs of CKM matrices for neutrinos:

\be
\overline{\chi^{(\nu)}_{+1i}}  [ (\U^{\nu}_{11})^T  
(\U^{l}_{11})^{\ast} ]_{ij} \gamma^{\mu} (\chi^{(l)}_{+1j} - \chi^{(l)'}_{-1j})
\ee

\be
\overline{\chi^{(\nu)}_{+2i}} [ (\U^{\nu}_{12})^T 
(\U^{l}_{11})^{\ast} ]_{ij} \gamma^{\mu} (\chi^{(l)}_{+1j} - \chi^{(l)'}_{-1j})
\ee

\be
-\overline{\chi^{(\nu)'}_{-1i}} [ (\U^{\nu}_{11})^T 
(\U^{l}_{11})^{\ast} ]_{ij} \gamma^{\mu} (\chi^{(l)}_{+1j} - \chi^{(l)'}_{-1j})
\ee

\be
-\overline{\chi^{(\nu)'}_{-2i}} [ (\U^{\nu}_{12})^T 
(\U^{l}_{11})^{\ast} ]_{ij} \gamma^{\mu} (\chi^{(l)}_{+1j} - \chi^{(l)'}_{-1j})
\ee

The essential feature is that all four distinct neutrino mass eigenstates
(per generation) are connected to the one combination of the lepton mass
eigenstates (which constitutes a lepton of definite handedness). The mixing
between the light neutrinos, $\chi_{\pm 1i}$, is given by $[ (\U^{\nu}_{11})^T  
(\U^{l}_{11})^{\ast} ]_{ij} $ while the mixing between the heavy neutrinos,
$\chi_{\pm 2i}$, is given by $[ (\U^{\nu}_{12})^T (\U^{l}_{11})^{\ast} ]_{ij}$. 

$\U^{\nu}_{11} \approx 1$ and $\U^{\nu}_{12} \approx 0$ 
in the limit of $\frac{M_D}{M_R} \ll 1$. 
Also, $\U^{\nu}_{11}$ and $\U^{\nu}_{12}$ are unitary only to zeroth order. 
First order correction spoil their unitarity and so the unitarity of the 
neutrino CKM matrices. 
The leptonic analogs of the CKM
matrix are thus two $3\times3$ \emph{nonunitary} matrices in generation space.
Even confining oneself to the weak interaction sector of 
the theory, this implies a much richer structure for CP violation
than had hitherto been anticipated. 
The mixing matrix (and hence the rich CP violating structure)
will also appear in other sectors of the theory,
in particular the interactions of the charged and neutral
leptons with the light Higgs particle(s).

In the pure Dirac case (quark case) $\U^{q}_{12}=0$, while
$\U^{q}_{11}$ is unitary, so, we have only one unitary, $3\times3$ CKM
matrix, which is the standard result.

\section{Conclusion}

With the recent data from the super-Kamiokande collaboration
reinforcing so strongly the case for non-zero neutrino masses,
the need for a comprehensive generic pedagogical understanding 
of neutrino masses and mixings is pressing.

We have shown that in the general case of complex  Dirac and Majorana
mass parameters, the neutrino ``mass matrix'' (i.e. the rest Hamiltonian) 
should be thought of as a $24\times24$ matrix (c.f. equation (\ref{eom})),
although a reduction to $12\times12$ is immediate.  A reduction
to a $6\times6$ mass matrix is possible but one must be cautious
in interpreting the results. In particular, the mass eigenvalues are correct,
but eigenstates are not.
In the usual limit of vanishing Majorana masses (quark case) the standard
treatment in terms of a $6\times6$ matrix can be readily recovered.  

We have shown that the mass eigenstates do obey a Majorana condition.

We have argued that the standard see-saw mechanism 
is preserved even in the case of complex mass matrices
so long as the scale of the sterile  (right-handed) Majorana masses
is large compared to that of the Dirac masses.

We have argued that in the limit $M_D/M_R\ll1$, the right handed
fields decouple from the problem. 

Finally, we have briefly hinted at the richness of the  problem of
mixing and CP violation in the leptonic sector.  
The group U(6) of $6\times6$ unitary matrices  has $36$
parameters. The mass matrix has $21$ magnitudes, and $21$ phases
of which $15$ phases are not removable by phase redefinitions. 
The leptonic CKM matrix is not a $3\times3$ unitary
matrix, so that it can contain more than the usual
one CP violating phase.

\vfil
\eject

Acknowledgements\hfil\break

The authors wish to acknowledge extensive conversations
with Cyrus Taylor and Mark Trodden, as well as many
useful conversations with Lawrence Krauss and Tanmay
Vachaspati during this work.  We thank in particular
Jim Cline for pointing out some useful decompositions of 
$H_0$.  GDS is supported by NSF CAREER.

\vspace{.5in}
\appendix{\bf{Appendix}}
\vspace{.2in}

The Lagrangian  density (\ref{2cL}) can be readily cast into the more 
familiar four-component form:
\be
\label{L}
-{\cal L} = i \bar{\psi_i} \gamma^{\mu} \partial_{\mu} \psi_i 
+ \left[ \ 
\overline{\psi_{iL}} \, M_{Dij} \, \psi_{jR} + 
+ \half \overline{(\psi_{iR})^c} \, M_{Rij} \, \psi_{jR} 
+ \half \overline{\psi_{iL}} \, M_{Lij} \, (\psi_{jL})^c 
+ \mathrm{h.c.} \ \right] 
\ee 
\noindent where $\psi$ is a four-component Dirac spinor, 
\be
\label{psi}
\psi = \left( \bav \nu_{a} \\ \Nbar^{\adot} \eav \right)
\ee
and we use the Weyl or spinor representation in which
\be
 \gamma^{\mu} = \left( \bam 0 & \sigma^{\mu} \\ \sigmabar^{\mu} & 0
\eam \right)
\ee
In this representation, the projectors onto the left and right subspaces are:
\be
 P_L = \half \, (\one + \gamma_{5} ) \ \ \ {\rm and} \ \ \ \ \ 
   P_R = \half \, (\one - \gamma_{5} )  \ \ \ \ \rm{where} \ \ \ \ 
   \gamma_{5} = \left( \bam \one & \ \ 0 \\ 0 & -\one \eam \right) 
\ee
Thus,
\be
 \psi_L = P_L \psi = \left( \bav \nu_{a} \\ 0 \eav \right)
\ \ \ \ \ \ \ \mathrm{and} \ \ \ \ \ \ \ \ 
\psi_R = P_R \psi = \left( \bav 0 \\ \Nbar^{\adot} \eav \right) 
\ee
describe left and right chirality neutrinos.
$\bar{\psi}$ is defined as:
\be
 \bar{\psi} = \psi^{\dagger} \gamma^{0} = (N^{a} \ \nubar_{\adot})\ \ ,
\ee
while charge conjugation is defined by the relation:
\be
 \psi^c = C \bar{\psi}^{T} \ \ \ \mathrm{where} \ \ \  
C = i \gamma^2 \gamma^{0} = \left( \bam i \sigma^2 & \ \ 0 \\
0 & i \sigmabar^2 \eam \right) 
\ee
\noindent Thus
\be
 (\psi_L )^c = \left( \bav 0 \\ \nubar^{\adot} \eav \right)
\ \ \ \ \ \ \mathrm{and} \ \ \ \ \ \ 
(\psi_R )^c = \left( \bav N_{a} \\ 0 \eav \right) 
\ee

\end{document}